# An all optical phase shifter and switch near 1550nm using tungsten disulfide (WS$_2$) deposited tapered fiber

**KAN WU,**[1,*] **CHAOSHI GUO,**[1] **HAO WANG,**[1] **XIAOYAN ZHANG,**[2] **JUN WANG,**[2] **AND JIANPING CHEN**[1]

[1]*State Key laboratory of Advanced Optical Communication Systems and Networks, Department of Electronic Engineering, Shanghai Jiao Tong University, Shanghai 200240, China*
[2]*Key Laboratory of Materials for High-power Laser, Shanghai Institute of Optics and Fine Mechanics, Chinese Academy of Sciences, Shanghai 201800, China*
*\*kanwu@sjtu.edu.cn*

**Abstract**: All optical phase shifters and switches play an important role for various all optical applications including all optical signal processing, sensing and communication. In this paper, we demonstrate a fiber all optical phase shifter using few-layer 2D material tungsten disulfide (WS$_2$) deposited on a tapered fiber. WS$_2$ absorbs injected 980 nm pump (control light) and generates heat which changes the refractive index of both WS$_2$ and tapered fiber due to thermo-optic effect and achieves a maximum phase shift of 6.1π near 1550 nm. The device has a loss of 3.7 dB. By constructing a Mach-Zehnder interferometer with WS$_2$ based phase shifter in one arm, an all optical switch is also obtained with an extinction ratio of 15 dB and a rise time of 7.3 ms. This all fiber low-cost and compact optical phase shifter and switch demonstrates the potential of 2D transition metal dichalcogenides for all optical signal processing devices.

## 1. Introduction

Two-dimensional (2D) materials have attracted intense attention after the discovery of graphene [1-4]. Many optical devices based on graphene has been demonstrated including optical modulator [4-6], photodetector [3], polarizer [7] and switch [8]. Besides graphene, novel 2D materials have been reported, including graphene oxide[9], topological insulators (TIs) [10-14], transition metal dichalcogenides (TMDs) [15-24], and black phosphorus (BP) [25-29], etc. Among these 2D materials, TMDs have been paid special attention because TMDs represent a class of materials with similar photonic and electronic properties, including high optical nonlinearity [30] and saturable absorption [17]. Investigation on one TMD material may benefit the research on other TMDs. Recently, a variety of TMDs have been demonstrated for mode-locked and Q-switched lasers as saturable absorbers [15-24].

Meanwhile, all optical devices at telecommunication wavelength of 1550 nm are of importance for modern optical communications such as all optical routing, modulation and optical logic gate. A compact and low-cost fiber device with such functions is always desired to be compatible to current fiber system. A few demonstrations of all optical modulator and phase shifter based on graphene have been reported [5, 6, 8]. Graphene has a uniform absorption in a wide spectrum which is good for absorption based devices. However for some other all optical devices based on optical phase tuning such as phase shifter, this is not desired because it induces additional loss at signal wavelength. For these phase-tuning devices the wideband absorption of graphene becomes a drawback and a better choice is the material that absorbs more at the wavelength of control light and less at the wavelength of signal light, i.e., the material should have non-uniform absorption in the spectrum. Moreover, the choice of control wavelength should be compatible to current fiber system near 1550 nm so that the device can be easily incorporated into telecommunication applications.

In this work, a proof-of-concept fiber all optical phase shifter operating near 1550 nm is fabricated by depositing few-layer TMD material tungsten disulfide ($WS_2$) to a tapered fiber. The bandgap of $WS_2$ near 1.3 eV (954 nm) [17] allows good absorption of the pump (control light) at 980 nm and weak absorption of the signal light near 1550 nm. The phase shifter is based on thermo-optic effect. When the pump at 980 nm is injected to the deposition area, $WS_2$ absorbs the light, generates heat and changes the refractive index of both $WS_2$ and tapered fiber and thus tunes the optical phase of 1550 nm signal light propagating through the fiber. The maximum phase tuning range is 6.1π with a slope efficiency of 0.0174π/mW. By incorporating this $WS_2$ based phase shifter to a fiber Mach-Zehnder interferometer (MZI), an all optical switch is also demonstrated. The switch has an extinction ratio of 15 dB and a rising time of 7.3 ms. Noting the deposition length of $WS_2$ is only ~500 μm, this multi-function device based on $WS_2$ deposited tapered fiber has the potential to be very compact for many all optical applications including optical routing, optical logic gate and optical signal processing.

## 2. Material preparation and characterization

The $WS_2$ deposited tapered fiber is the key element in the all optical phase shifter. It is fabricated using the following two steps. Firstly, $WS_2$ nanosheets are prepared using standard liquid-phase exfoliation (LPE) method [18]. Briefly, 5 mg $WS_2$ powders are dispersed in 1.5 mg/ml sodium cholate (SC, as surfactant) aqueous solution and sonicated for 1 hour using a horn probe sonic tip. The dispersions are then centrifuged at 3000 rpm for 90 minutes to remove the unexfoliated flakes. The top 1/2 of the dispersions is collected by pipette. The prepared $WS_2$ dispersions have a concentration of ~0.015 mg/ml. Figure 1(a) shows the transmission electron microscope (TEM) image of the prepared $WS_2$ nanosheets. The typical size of nanosheets is a few hundreds of nanometers. Secondly, optical driven deposition is utilized to deposit $WS_2$ nanosheets to the tapered fiber [31]. Briefly, a tapered fiber (SMF28) with a diameter of ~10 μm is immersed in $WS_2$ dispersions. By applying light through the tapered fiber, leaked optical field near the tapered fiber attracts the $WS_2$ nanosheets to the tapered fiber. The output power of the tapered fiber is monitored by an optical power meter to control the deposition process. The loss of the fabricated $WS_2$ tapered fiber is 3.5 dB near 1550 nm. In the whole fabrication, there is no other chemicals or index matching gel used.

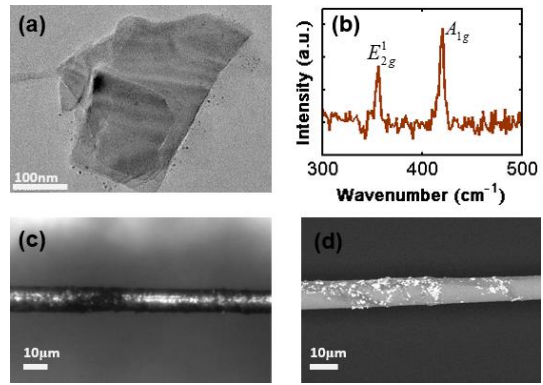

Fig. 1. (a) TEM image of $WS_2$ nanosheets in the dispersions, (b) Raman spectrum of $WS_2$ deposited on tapered fiber, (c) microscopic and (d) SEM images of $WS_2$ deposited tapered fiber

Raman spectroscopy is utilized to confirm the deposited material on tapered fiber, shown in Fig. 1(b). Clear fingerprint peaks of $WS_2$ can be observed. The distance between $E^1_{2g}$ and $A_{1g}$ peaks is 64.3 cm$^{-1}$, meaning the average number of layers is 3-5 [32]. Microscopic image of WS$_2$ tapered fiber is shown in Fig. 1(c). Deposited material can be clearly observed. The length of the deposited region is ~500 μm. Because scanning electron microscope (SEM) image requires gold deposition on the target sample which will damage our prepared WS$_2$ tapered fiber. A similar sample of WS$_2$ tapered fiber is used for the SEM imaging, shown in Fig. 1(d). It should be noted that in an optical driven deposition process WS$_2$ nanosheets are randomly attached to the waist region of the tapered fiber, which is different from the graphene deposition based on film transfer.

## 3. Experiments and results

### 3.1 All optical phase shifter

The all optical phase shifter is shown in the red frame in Fig. 2. It consists of two 980/1550 wavelength demultiplexers (WDMs) and a WS$_2$ deposited tapered fiber in the middle. 980 nm pump (control light) is injected from the first WDM and the residual 980 nm light is extracted by the second WDM. WS$_2$ absorbs 980 nm light, generates heat and changes the refractive index of both WS$_2$ and tapered fiber due to thermal-optic effect. When 1550 nm signal light propagates through the tapered fiber, its optical phase is modified accordingly. The index temperature coefficient of WS$_2$ is estimated to be $2.48\sim5.48\cdot10^{-4}$/°C. More details are given in section 4.2 in Discussion. Although 980 nm light is slightly below the bandgap of ~1.3 eV (954 nm) for few-layer WS$_2$, edge states of WS$_2$ nanosheets has been found to allow the absorption to extend to near 2 μm [33]. Moreover the absorption of WS$_2$ decreases with the increase of wavelength which allows higher absorption at pump wavelength for better control efficiency and weaker absorption at signal wavelength for lower loss. The total loss of the phase shifter is ~3.7 dB near 1550 nm (0.2 dB from WDMs) and ~6 dB at 980 nm (1 dB from WDMs). To measure the phase shift value, the phase shifter is embedded into one arm of a fiber Mach-Zehnder interferometer, shown in Fig. 2. A sweeping continuous wave (CW) laser source near 1550 nm (Agilent 81640A) is used to measure the shift of the transmission spectrum. A 70:30 fiber splitter is used and 70% of the input 1550 nm light power is fed into the phase shifter's arm to compensate the loss. In the other arm of the MZI, a tunable delay is inserted to better adjust the arm length. Two arms are finally combined together by a 50:50 fiber coupler. The output from the coupler is monitored by an optical spectrum analyzer (Yokogawa AQ6370C) for its spectral properties, and a 2.5 GHz oscilloscope (Agilent DSO9254A) with a 10 GHz photodetector (EOT 3500F) for its time-domain properties. The fiber lengths of two arms are very close to each other to guarantee a good interference result.

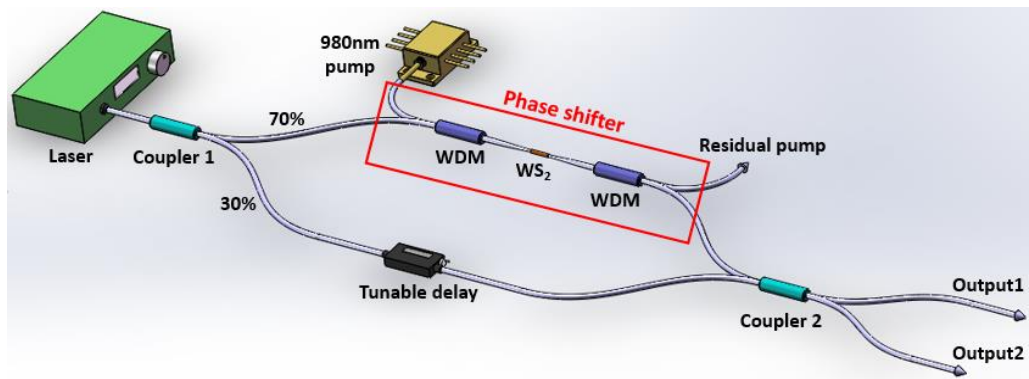

Fig. 2. Experimental setup of the all optical phase shifter and switch based on WS$_2$ deposited tapered fiber.

Figure 3 summarizes the phase shift property of the device. The blue curve in Fig. 3(a) shows the transmission spectrum at 0 phase shift without 980 nm pump. The free spectral range (FSR) is 0.173 nm, corresponding to a frequency FSR of 21.73 GHz near 1565 nm. The time difference of two arms is therefore 46.24 ps and the length difference is 9.43 mm. When the 980 nm pump is injected, $WS_2$ absorbs the pump light, generates heat and modifies the refractive index of both $WS_2$ and tapered fiber due to thermo-optic effect. The transmission loss of 980 nm pump from the first WDM to the second WDM is ~6 dB with 1 dB loss from two WDMs and 5 dB absorption from the $WS_2$ tapered fiber. When 980 nm pump is increased to 350 mW, the spectrum shifts ~0.53 nm to the right side, corresponding to a phase shift of ~6.1π. A typical spectrum with 5π phase shift (spectral shift of 2.5 FSRs) is shown as red curve in Fig. 3(a). In both two spectra, an extinction ratio of more than 15 dB can be observed.

For a thermal-optic device, it is the average power rather than the peak power that works. So we then change the light source to a mode-locked laser (1560 nm center wavelength, 20 nm bandwidth, 37 MHz repetition rate) with low average power (<4 mW) to monitor the spectral shift in real time when pump is applied. The measured transmission spectra are nearly identical to the results by using a CW laser, shown in Fig. 3(b). Although $WS_2$ tapered fiber exhibits saturable absorption with a modulation depth of ~3% (0.15 dB), this contribution on the intensity is much weaker than the intensity variation of 15 dB caused by the MZI. In the supplementary video (see Visualization 1), we apply a saw wave 980 nm pump light (power changing from 0 to 250 mW, 3 s period) to the phase shifter, the rightwards and leftwards movement of the spectrum can be clearly observed.

The phase shift with respect to the pump power is shown in Fig. 3(c). Mode-locked laser source is used. Two different input average power levels of 5 dBm and -5 dBm are measured with nearly identical results, indicating that the phase shifter is less sensitive to the input power. For both power levels, a nearly linear relation is observed with a slope efficiency of ~0.0174π/mW.

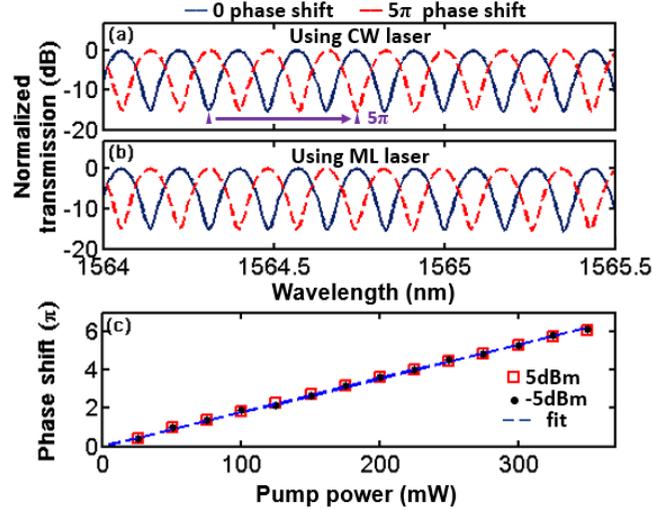

Fig. 3. Transmission spectra of the MZI with 0 (blue) and 5π phase shift (red) using (a) a continuous wave laser source and (b) a mode-locked laser source (see Visualization 1). (c) Phase shift with respect to the injected pump power with two input power levels.

There are a few reasons of choosing 980 nm wavelength as the pumping wavelength. Firstly, $WS_2$ has a bandgap of ~1.96 eV for monolayer and ~1.3 eV for bulk [17], corresponding to the wavelength of 632 nm and 954 nm. Edge states in the $WS_2$ nanosheets can further extend the absorption spectrum of $WS_2$ and other TMDs to longer wavelength of 2 μm [33], but the absorption decreases with the increase of wavelength. Therefore pumping wavelength should be close to short wavelength for higher absorption. An obvious experimental support is that the insertion loss of $WS_2$ tapered fiber is 3.5 dB at 1550 nm and 5 dB at 980 nm. Secondly, the pump light should be easy to combine with 1550 nm light for an all fiber system. A fiber 980/1550 WDM is a very mature commercial product because 980 nm wavelength is also the pumping wavelength for erbium-doped fiber (EDF) in fiber laser and fiber amplifier near 1550 nm. Moreover, the availability of high power 980 nm pump laser diode can provide much flexibility in the experiment.

*3.2 All optical switch*

The whole MZI can also be utilized as an all optical switch with two outputs. In the following experiments, a CW laser source centered at 1560 nm is used. A square wave electronic signal is injected into the analog modulation port of the pump driver and the pulsed 980 nm pump light is generated. Pump light is then applied to the phase shifter. Figure 4(a) shows the measured waveforms of pump power (yellow) and switch output at "output 1" port (blue). The repetition rate of the square wave is 10 Hz with a 30% duty cycle (i.e., 30 ms on, 70 ms off). A clear on-off operation of the switch is realized. The on-off extinction ratio is 15 dB obtained from Fig. 3(a) and 3(b). Bandwidth limitation effect can also be observed that both the rising edge and the falling edge of the switch output have been flattened. A zoomed view of a single off-on-off transition of the switch output is shown in Fig. 4(b). By using exponential fit $(1-\exp(-t/t_r))$ for the rising edge and $\exp(-t/t_f)$ for the falling edge, the time constants are found to be $t_r$=7.3 ms for the rising edge and $t_f$=3.5 ms for the falling edge, shown as red curve in Fig. 4(b). The time constant for the rising edge is related to the process that $WS_2$ absorbs pump light to raise the temperature and meanwhile heat gradually dissipates due to the air convection. The time constant for the falling edge is purely related to the heat dissipation of $WS_2$ and the tapered fiber because there is no pump light. Fig. 4(c) shows the complementary output at "output 2" port.

It is well known that the transmission curve of an MZI is a cosine function $(1+\cos\Delta\varphi)/2$ where $\Delta\varphi$ represents the phase difference in the two arms of the MZI. If the peak power of the applied pump light is further increased to exceed $\pi$ phase shift of the MZI, breaking of the output waveform from the switch can be observed, shown in Fig. 4(d). In this experiment, the peak power of the pump light is doubled. When the instant power of pump light increases at the beginning, the instant power of the switch output also increases, same as the normal case in Fig. 4(a). However, when the instant power of the pump light keeps increasing, the phase difference $\Delta\varphi$ exceeds $2\pi$ and the MZI becomes biased reversely. Then the instant power of switch output decreases with the increase of instant power of pump light. When the instant power of pump light becomes to drop, all the above processes are reversed. Therefore one input square wave pump pulse breaks to two output pulses with one valley in the middle. This is similar to the case that the driving voltage exceeds $V_\pi$ in an MZI based electro-optic modulator. One can also note that the rising edge of such over-driven switch output is much sharper than that of normally driven output in Fig. 4(a). This is due to the fact that when the maximum pulse peak power of the pump light is increased, $WS_2$ absorbs more light in the same time and generates more heat. As a result, the refractive index changes faster. This may help to increase the turn-on speed of the switch if the input signal is short enough to pass the switch before the switch transmission drops.

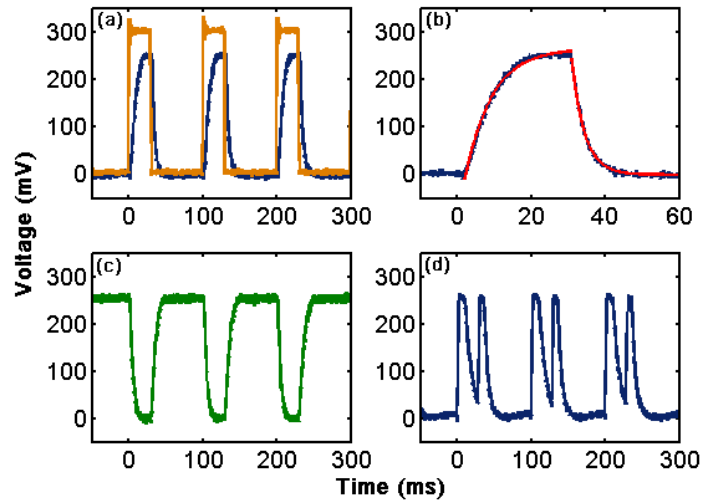

Fig. 4. (a) Pulsed pump light (yellow) and switch output at "output 1" port (blue), (b) a zoomed view of a single off-on-off transition of the switch (blue) and exponential fit (red), (c) complementary output at "output 2" port, and (d) output breaking when the MZI is over-driven.

To further investigate the response speed of the phase shifter and switch, a sine wave waveform of the pump power is applied to measure the bandwidth of the device. The frequency of the sine-wave pump is varied from 10 Hz to 340 Hz by injecting electronic signal into the analog modulation port of the pump driver. The bandwidth of the analog modulation port is more than 1 kHz. When this sine wave pump light is applied to the phase shifter and switch, the output power from the MZI also exhibits sine wave waveform. The output waveform is measured by a 10 GHz photodetector and an oscilloscope. By comparing the amplitude of the output sine wave signal at different frequencies, the bandwidth of the device can be estimated. Typical sine wave power waveforms for the pump and the switch output are shown in Fig. 5(a) and 5(b). The frequency response of the phase shifter and switch is shown in Fig. 5(c) when the modulation frequency varies from 10 Hz to 340 Hz. It can be found that the 3-dB bandwidth of the device is near 320 Hz.

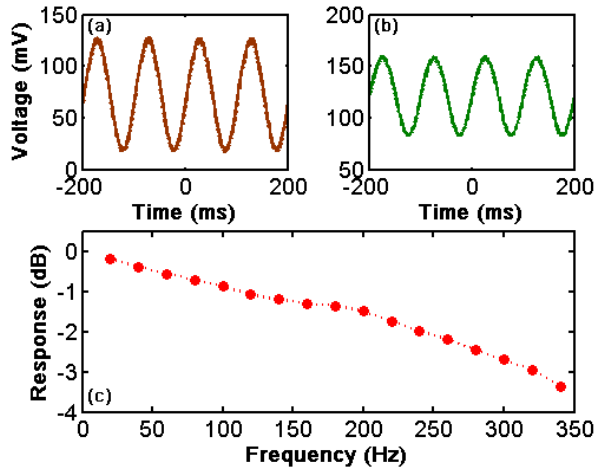

Fig. 5. Sine wave waveforms of (a) pump and (b) switch output with a repetition rate of 10 Hz, and (c) frequency response of the switch

## 4. Discussion

### 4.1. Comparison among different all optical devices

In this work, our WS$_2$ based all optical phase shifter and switch is aimed for all optical signal processing. Our purpose is to provide a fiber compatible, low-cost, simple-fabrication and compact solution for all optical switches. Meanwhile there are many applications of optical switches that do not require high speed. For example, the connections among different fiber links in the data center are usually re-allocated in a frequency of hours or days and a switching time of a few milliseconds is acceptable.

Here we provide comparisons with different reported all optical switches. The comparison includes two parts: comparison between 2D materials and other nonlinear materials which shows the advantages brought by the unique properties of 2D materials, and comparison between WS$_2$ based and graphene based switches which explains the motivation of using WS$_2$.

Table 1 summarizes the comparison among 2D materials and other nonlinear materials. For all optical switches, the key components are the nonlinear materials. For the designs with highly nonlinear fiber (HNLF), the systems usually require meter-long HNLF. The system cannot be compact. The cost of HNLF is also high. For the designs with nonlinear waveguide, such as silicon and chalcogenides, the systems can be compact because of the small chip size, but the coupling between fiber system and chip usually leads to at least 3-5 dB loss. Moreover, the waveguide itself adds additional loss due to the unsmooth side wall and free carrier absorption (for semiconductor materials). The fabrication cost is also high due to many steps of fabrication in the clean room. In contrast, 2D materials based optical switches can be potentially very compact because the light-material interaction region is only micron to millimeter long. The loss is not very high because the deposition/transfer process can be controlled. The cost is also low because no complicated fabrication is involved.

Table 1. Comparison among all optical devices based on 2D materials and other nonlinear materials

|  | *Compact* | *Loss* | *Cost* | *Fabrication* |
|---|---|---|---|---|
| **HNLF** | No | Low | High | Complicated |
| **Nonlinear waveguide** | Yes | High | High | Complicated |
| **2D materials** | Yes | Medium | Low | Simple |

Table 2 summarizes the comparison between WS$_2$ based and graphene based all optical switches. Both WS$_2$ and graphene belong to 2D materials, however, there are still a few different properties when applied to an all optical switch. The comparison is performed with a recently published work on a graphene based all optical switch [8], which presented a very impressive work and partially triggered our idea. We fully respect authors' work in Ref. [8].

Graphene in Ref. [8] is fabricated by chemical vapor deposition (CVD) and transferred to the tapered fiber by etching away the copper foil. The graphene is very thin (a few nanometers) and therefore the light-material interaction with graphene on the cross section of tapered fiber is weak. In our work, WS$_2$ nanosheets are fabricated by liquid phase exfoliation (LPE) and transferred to the tapered fiber by optical driven deposition. The transfer process can be monitored in real time and the total loss of WS$_2$ deposited tapered fiber can be controlled by controlling the deposition time. The thickness of WS$_2$ surrounding the tapered fiber is micrometer level, which enhances the light-material interaction on the cross section. Therefore, in our device, the interaction length is very short, only 500 μm, showing the great potential for a very compact device. Moreover, the interaction between 980 nm light and material surrounding the tapered fiber is naturally weaker than 1550 nm because the beam size of 980 nm is smaller than that of 1550 nm in the tapered fiber. If a material with uniform absorption at 980 nm and 1550 nm is used, such as graphene, the device absorption at 980 nm would be smaller than that at 1550 nm, which obviously reduces the control efficiency of the device. In our work, we use $WS_2$ which has a higher absorption at 980 nm and compensates the weaker light-material interaction at 980 nm caused by the smaller beam size. In table 2, it can be found that our device has a better control efficiency which is defined as the phase shift normalized to pump power and interaction length, or slope efficiency (=phase shift/pump power) divided by interaction length.

Table 2. Comparison between our work and graphene based device

|  | Material thickness on tapered fiber | Light-material interaction on cross section | Interaction length | Control efficiency | Loss @1550 | Loss @980 |
|---|---|---|---|---|---|---|
| $WS_2$ based device | μm level | Strong | 500 μm | 0.035π/mW/mm | 3.5 dB | 5 dB |
| Graphene based device | nm level | Weak | 5 mm | 0.018π/mW/mm | 5.4 dB | 2.1 dB |

### 4.2. Index temperature coefficient of $WS_2$

The index temperature coefficient of $WS_2$ can be roughly estimated. The mode distribution on the cross section of $WS_2$ tapered fiber is shown in Fig. 6. The thickness of deposited $WS_2$ is estimated to be 0.5 μm and the diameter of tapered fiber is 10 μm. The refractive index of $WS_2$ is estimated to be ~3.67 at 980 nm and ~3.20 at 1550 nm by extending the data in Ref. [34] to the near IR region. The power percentage is ~3% in the $WS_2$ area and ~96% in the tapered fiber for both 980 nm and 1550 nm because the index of $WS_2$ is higher at 980 nm than that at 1550 nm.

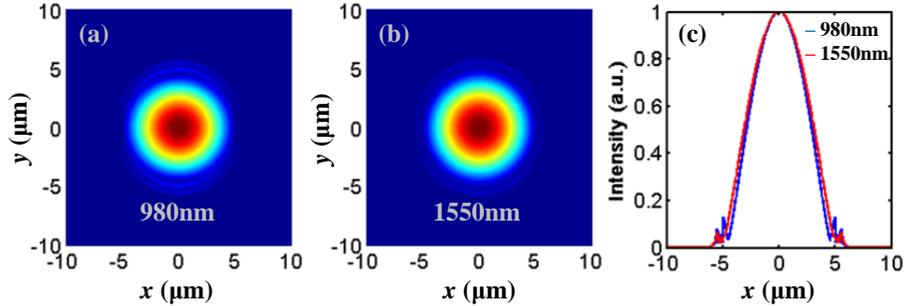

Fig.6. Mode distribution of (a) 980 nm light and (b) 1550 nm light, and (c) mode distribution in the middle line

By comparing with the data in a similar graphene based all optical switch [8], we roughly estimate the local temperature change of $WS_2$ tapered fiber is 40~60°C for a π phase shift. Using the relation of $\Delta n \cdot L = \lambda/(2n)$, where the interaction length L = 500 μm, operating wavelength λ = 1550 nm and effective index n = ~1.44. The index change of our device is $\Delta n = 1.08 \cdot 10^{-3}$. Because the power percentage in $WS_2$ and tapered fiber is ~3% and 96%, the contribution of index change is estimated by $\Delta n = \Delta n_{WS2} \cdot 3\% + \Delta n_{TF} \cdot 96\%$. Noting that the index temperature coefficient for silica fiber is ~$1.1 \cdot 10^{-5}$/°C, the index change of tapered fiber $\Delta n_{TF}$ is $4.4 \sim 6.6 \cdot 10^{-4}$ for a temperature change of 40~60°C. The index change of $WS_2$ $\Delta n_{WS2}$ is then found to be $1.49 \sim 2.19 \cdot 10^{-2}$, corresponding to an index temperature coefficient of $2.48 \sim 5.48 \cdot 10^{-4}$/°C. This result indicates that $WS_2$ and tapered fiber nearly have equal contribution on the index change.

## 5. Conclusions

In conclusion, we have demonstrated a fiber all optical phase shifter and switch based on $WS_2$ deposited tapered fiber. By absorbing the pump (control light) at 980 nm, $WS_2$ generates heat and changes the refractive index of both $WS_2$ and tapered fiber due to thermo-optic effect. For the phase shifter, a phase shift of 6.1π is obtained under a maximum pump power of 350 mW, corresponding to a slope efficiency of 0.0174π/mW. For the switch, the on-off extinction ratio of ~15 dB is obtained with a rise time of 7.3 ms and a 3-dB bandwidth of 320 Hz. Our work may benefit the research on photonic applications of 2D TMD materials and pave the way of TMD materials for practical applications such as all optical signal processing and optical routing.

## Funding

This work is partially supported by NSFC (No. 61505105, No. 51302285, No. 61522510), Shanghai Yangfan Program (No. 14YF1401600), the External Cooperation Program of BIC, CAS (No. 181231KYSB20130007), the Fundamental Research Funds for the Central Universities (No. 2232014D3-28), the "Strategic Priority Research Program" of CAS (No. XDB160307), and Open Fund of IPOC (BUPT).